\documentclass[useAMS,usegraphicx]{mn2e}
\usepackage{epsfig}
\newcommand{\src} {SMM2}
\newcommand{\srce} {SMM2E}

\newcommand{\mum}   {$\mu$m}
\newcommand{\kms}   {km~s$^{-1}$}

\newcommand{\jpb}   {$\rm Jy~beam^{-1}$}    
\newcommand{\lo}    {$L_{\sun}$}
\newcommand{\mo}    {$M_{\sun}$}
\newcommand{\ro}    {$R_{\sun}$}
\newcommand{\mj}    {$M_\mathrm{Jup}$}
\newcommand{\co}    {$^{12}$CO}
\newcommand{\tco}    {$^{13}$CO}
\newcommand{\ceo}    {C$^{18}$O}

\newcommand{\et}    {et al.}
\newcommand{\eg}    {e.\,g.,}

\newcommand{\supa}  {$^\mathrm{a}$}
\newcommand{\supb}  {$^\mathrm{b}$}
\newcommand{\supc}  {$^\mathrm{c}$}

\newcommand{\phb}   {\phantom{$>$}}

\title[A proto-BD driving an outflow]{IC\,348-SMM2E: a Class 0 proto-brown dwarf candidate forming as a scaled-down version of low-mass stars}
\author[Palau et al.]{Aina Palau$^{1}$\thanks{E-mail:a.palau@crya.unam.mx}, 
Luis A. Zapata$^{1}$, Luis F. Rodr\'iguez$^{1}$, Herv\'e Bouy$^2$, David Barrado$^2$,
\newauthor
Mar\'ia Morales-Calder\'on$^2$, Philip C. Myers$^3$, Nicholas Chapman$^4$, Carmen Ju\'arez$^5$,
\newauthor
Di Li$^{6, 7}$
\\
$^{1}$ Centro de Radioastronom\'ia y Astrof\'isica, Universidad Nacional Aut\'onoma de M\'exico, P.O. Box 3-72, 58090 Morelia, Michoac\'an, M\'exico\\
$^{2}$ Centro de Astrobiolog\'ia, INTA-CSIC, Depto. Astrof\'isica , ESAC Campus, P.O. Box 78, 28691 Villanueva de la Ca\~nada,
Madrid, Spain\\
$^{3}$ Harvard-Smithsonian Center for Astrophysics, 60 Garden Street, Cambridge, MA 02138, USA\\
$^{4}$ Center for Interdisciplinary Exploration and Research in Astrophysics (CIERA) and Department of Physics \&
Astronomy, Northwestern\\ 
University, 2145 Sheridan Road, Evanston, IL 60208, USA\\
$^{5}$ Institut de Ci\`encies de l'Espai (CSIC-IEEC), Campus UAB-Facultat de Ci\`encies, Torre C5-parell 2, 08193 Bellaterra, Catalunya, Spain\\
$^{6}$ National Astronomical Observatories, Chinese Academy of Sciences, Beijing 100012, China\\
$^{7}$ Space Science Institute, Boulder, CO. 80301, USA\\
}
\begin{document}

\date{Accepted date. Received date; in original form date}

\pagerange{\pageref{firstpage}--\pageref{lastpage}} \pubyear{2012}

\maketitle

\label{firstpage}

\begin{abstract}
We report on Submillimeter Array observations of the 870~\mum\ continuum and CO\,(3--2), \tco\,(2--1) and \ceo\,(2--1) line emission of a faint object, \srce, near the driving source of the HH\,797 outflow in the IC\,348 cluster. The continuum emission shows an unresolved source for which we estimate a mass of gas and dust of 30~\mj, and the CO\,(3--2) line reveals a compact bipolar outflow centred on \srce, and barely seen also in \tco\,(2--1). 
In addition, \ceo\,(2--1) emission reveals hints of a possible rotating envelope/disk perpendicular to the outflow, for which we infer a dynamical mass of $\sim16$~\mj.
In order to further constrain the accreted mass of the object, we gathered data from \emph{Spitzer, Herschel}, and new and archive submillimetre observations, and built the Spectral Energy Distribution (SED). The SED can be fitted with one single modified black-body from 70~\mum\ down to 2.1~cm, using a dust temperature of $\sim24$~K, a dust emissivity index of 0.8, and an envelope mass of $\sim35$~\mj. The bolometric luminosity is 0.10~\lo, and the bolometric temperature is 35~K. 
Thus, \srce\ is comparable to the known Class 0 objects in the stellar domain. An estimate of the final mass indicates that \srce\ will most likely remain substellar, and the \srce\ outflow force matches the trend with luminosity known for young stellar objects. 
Thus, \srce\ constitutes an excellent example of a Class 0 proto-brown dwarf candidate which forms as a scaled-down version of low-mass stars. 
Finally, \srce\ seems to be part of a wide ($\sim2400$~AU) multiple system of Class 0 sources.
\end{abstract}


\section{Introduction}\label{sintro}

The formation of brown-dwarfs (BDs) remains a highly debated field of current astrophysics. BD masses ($<0.075$~\mo) are much smaller than Jeans masses estimated for typical conditions of molecular clouds ($\sim1$~\mo), and their formation cannot be simply explained as a scaled-down version of low-mass stars. Several mechanisms have been proposed to solve this problem, such as ejection from fragmented massive disks (\eg\ Rice et al. 2003; Stamatellos \& Whitworth 2009; Basu \& Vorobyov 2012) or from multiple systems (Reipurth \& Clarke 2001; Bate, Bonnell \& Broom 2002; Umbreit et al. 2005), photo-erosion by nearby massive stars (\eg\ Hester et al. 1996; Whitworth \& Zinnecker 2004), and the formation of cores of very low Jeans masses through gravoturbulent fragmentation (\eg\ Padoan \& Nordlund 2004; Hennebelle \& Chabrier 2008; Chabrier et al. 2014). A method to distinguish which of these mechanisms dominates is to study BDs in their most embedded stages of their formation, comparable to the Class 0/I stages of low-mass star formation (\eg\ Andr\'e, Ward-Thompson, \& Barsony 1993), called here `proto-BDs'. If BDs form as low-mass stars, one would expect substantial envelopes and outflows as observed in young stellar objects.

With the great sensitivity of the {\it Spitzer Space Telescope}, a new family of objects emerged, the so-called Very Low Luminosity Objects (VeLLOs), which are young stellar objects in a quiescent accretion phase, or potential proto-BD candidates and very low-mass stars. VeLLOs are most of them associated with infrared sources and present internal luminosities $\la0.1$~\lo\ (\eg\ Young et al. 2004; di Francesco et al. 2007; Dunham et al. 2008; Hsieh \& Lai 2013). However, recent attempts to search for molecular outflows have reported a striking low detection rate (Schwarz et al. 2012), probably because their low-velocity wings cannot be separated from the large-scale ambient cloud emission detected by single-dish telescopes.
Thus, most of the VeLLOs need to be studied with millimetre interferometry, making difficult to infer general properties in a statistically significant sample. Another group of objects characterized by very low luminosities are the so-called First Hydrostatic Cores (FHCs). FHCs were theoretically predicted by Larson (1969) and are supposed to form during the first stages of collapse, once the density is large enough to turn the collapse from isothermal to adiabatic, providing the required pressure to balance gravity. The predicted properties of FHCs are low internal luminosities (with bolometric luminosities $<0.5$~\lo), very low masses (0.04--0.1~\mo, \eg\ Boss \& Yorke 1995; Saigo \& Tomisaka 2006),  Spectral Energy Distributions (SEDs) peaking around 100~\mum\ (\eg\ Omukai 2007) and association with low-velocity outflows (Machida et al. 2008). A number of FHC candidates actually present these properties (\eg\ Onishi, Mizuno \& Fukui 1999; Belloche et al. 2006; Enoch et al. 2010; Dunham et al. 2011; Pineda et al. 2011; Chen et al. 2010, 2012; Pezzuto et al. 2012; Murillo \& Lai 2013; Hirano \& Liu 2014). 
Thus, the properties of FHCs are similar to the properties expected for a deeply embedded Class 0-like proto-BD; it is the mass reservoir in the envelope (much larger for FHCs than for proto-BDs) that is the key distinguishing parameter between the two.

The number of proto-BD candidates proposed so far is very small. Barrado \et\ (2009) and Palau et al. (2012) present a Class 0/I candidate of 0.003~\lo,  J041757B, associated with a possible thermal radiojet but with no molecular outflow detected so far (Phan-Bao et al. 2014). A more luminous Class 0/I candidate driving a compact outflow is L1148-IRS (Kauffmann et al. 2011), with 0.12~\lo\ and an envelope mass of 0.14~\mo. Very recently, Lee et al. (2009, 2013) present a Class 0 object of 0.15~\lo, driving an outflow and with an even smaller envelope mass, 0.09~\mo, being thus the best proto-BD candidate known to date. With such a small number of proto-BD candidates, a primary urgent task is searching for further candidates in order to build a small sample and study whether their overall properties match the known relations of young stellar objects. In this work we present a new Class 0 proto-BD candidate driving a compact outflow and with only 0.03~\mo\ of envelope mass.

IC348-SMM2 is a submillimeter clump in the Perseus cloud at 240~pc of distance (Hirota et al. 2008, 2011; Chen et al. 2013). The clump, detected by Walawender et al. (2006) using the James Clerk Maxwell Telescope (JCMT), is also detected with the Caltech Submillimeter Observatory (CSO) at 350~\mum\ (Davidson et al. 2011) and \emph{Herschel} (Sadavoy et al. 2014). At 1.3~mm, interferometric observations resolve the IC\,348-SMM2 clump into two sources, a strong source driving a well-developed north-south outflow, also known as HH\,797 (Pech et al. 2012; Chen et al. 2013), and a fainter object about 10~arcsec to the (north)east, referred to as `\srce'. 
Here we report on Submillimeter Array (SMA) observations which clearly show association of molecular gas with \srce, providing evidence that it belongs to IC\,348. In addition, we complement the interferometric submillimetre data with new CSO observations and multiwavelength archive data that allow to gain insight into the nature of \srce.

\section{Observations}\label{sobs}

\subsection{SMA submillimetre data}

The 345~GHz archive SMA (Ho et al. 2004) observations of the HH\,797 outflow were carried out on 2006 Sep 8th in compact configuration, and with projected baselines ranging from 13 to 80~k$\lambda$.
The phase reference centre for the observations was at RA(J2000.0) = 03:43:57.100, Dec(J2000.0) = +32:03:04.80. 
A mosaic of 7 pointings separated 17~arcsec in Dec was performed to cover a total field of view of roughly $3\times0.6$~arcmin$^2$ oriented in the north-south direction and covering the strongest emission of the outflow. 
Receivers were tuned to cover the frequency range 345.589--347.589~GHz in the upper sideband (thus including the CO\,(3--2) line at 345.79599~GHz), and  335.589--337.589~GHz in the lower sideband. Channel spacing was set to 0.41~MHz (256 channels per chunk) across all the band, corresponding to a velocity resolution of 0.352~\kms. This was smoothed to 0.704~\kms. 
The FWHM of the primary beam at the frequency of observations is 37~arcsec.

Calibration was performed using the {\sc IDL} superset {\sc MIR} package\footnote{The MIR-IDL cookbook by C. Qi can be found at https://www.cfa.harvard.edu/$\sim$cqi/mircook.html.} (Scoville et al. 1993) and following the standard procedures.
Passband response was obtained from observations of PKS\,B1921$-$293, and 3C84 was used as a gain calibrator, for which typical rms of the phases were 55 per cent. The absolute flux density scale was determined from Uranus, and its uncertainty is estimated to be of 15--20 per cent.
The positional accuracy is estimated to be $\sim 0.3$~arcsec. 

Imaging was performed following the standard procedures in {\sc miriad} (Sault et al. 1995) and {\sc karma} (Gooch 1996). A {\sc robust} (Briggs 1995) parameter equal to 2 was used to obtain a good sensitivity. The mosaic was corrected for the primary beam attenuation, with noise decreasing at the edges of the field of view. The resulting synthesized beam and rms noise of the continuum image are $2.81\times2.32$~arcsec$^2$, P.A.$=87.37$\degr, and 17~m\jpb, and for the line we obtained a beam of $2.82\times2.32$~arcsec$^2$, P.A.$=87.38$\degr, and a rms noise of 0.8~\jpb\ per channel. 

In order to further study the CO\,(2--1), \tco\,(2--1) and \ceo\,(2--1) emission of \srce, we also imaged these transitions from the 230~GHz dataset published by Pech et al. (2012).
For CO\,(2--1), we selected a $uv$-range of 18--55~k$\lambda$, and used a robust parameter equal to $-2$, yielding a final rms noise of 0.12~\jpb\ per channel of 1.05~\kms\ width, and a synthesized beam of $3.14\times2.51$~arcsec$^2$, P.A.$=86.05$\degr.
For \tco\,(2--1) and \ceo\,(2--1), we used a robust parameter equal to $0$ (including the entire $uv$-range), yielding a final rms noise of 0.25~\jpb\ per channel of 0.28~\kms\ width, and a synthesized beam of $3.64\times3.04$~arcsec$^2$, P.A.$=83.44$\degr, for \tco, and $3.56\times3.06$~arcsec$^2$, P.A.$=87.37$\degr, for \ceo.

\subsection{CSO submillimetre data}

IC\,348-SMM2 was observed at the CSO in 2013 Nov 28 and 29 using the polarimeter for the Submillimeter High Angular Resolution Camera (SHARC-II, Dowell et al. 2003). SHARC-II is a camera of $32\times12$~pixels observing at 350~\mum, which can be used for polarization by placing a half-wave plate which is rotated at four different angles to determine the total flux and the linear polarization (Li et al. 2008). Although observations were carried out using the polarimeter, in this work we focus on the intensity image only of IC\,348-SMM2.
%
Observations were done in chop-nod mode, where an observation is made at each of the four half-wave plate angles, constituting one cycle.
The target was observed for 27 half wave plate cycles (a single cycle takes about 7 minutes).

Opacities at 225~GHz were around 0.05 for both days. 
Neptune was observed for initial focus and pointing, and CRL\,618 served for regular pointing corrections.
The absolute flux scale was determined from L1551, for which we adopted a peak intensity of 45.2~\jpb. The uncertainty in the absolute flux scale is estimated to be $\sim30$ per cent.
The final map size and beam are $1.5\times1.5$~arcmin$^2$ and 10~arcsec (FWHM), respectively. 
We followed the data reduction procedure as discussed in Davidson et al. (2011) and Chapman et al. (2013),
achieving a rms noise near the map centre of 0.6~\jpb.

\begin{figure*}
\begin{center}
\begin{tabular}[b]{cc}
    \epsfig{file=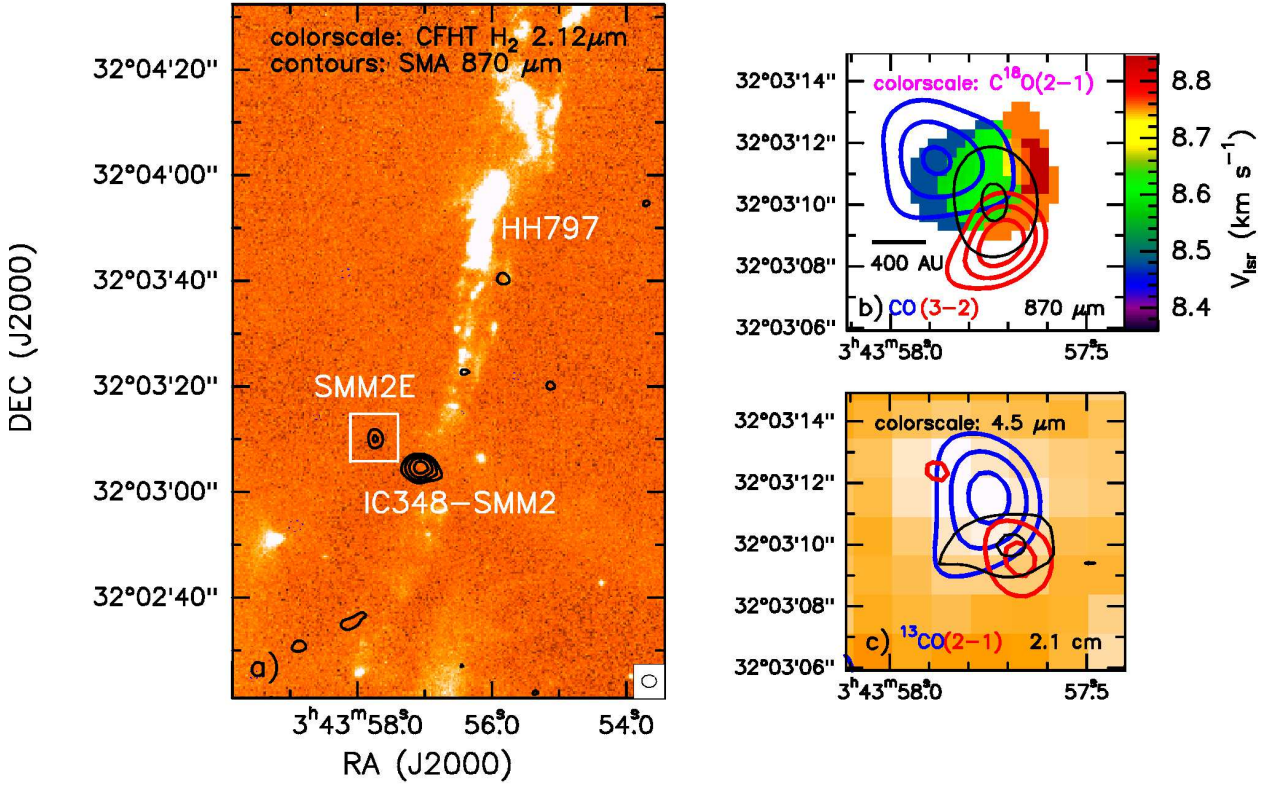, width=17cm,angle=0}\\
\end{tabular}
\caption{{\bf a)} Large-scale field of view where IC348-SMM2 is found.
Colorscale: CFHT H$_2$ archive image (Section~2.3.2).
Black contours: SMA 870~\mum\ continuum emission with contours $-3$, 3, 6, 12, and 24 times the rms noise of the map, 17~m\jpb.
{\bf b)} Zoom in on \srce: SMA CO\,(3--2) blueshifted and redshifted emission (blue, red contours) and SMA 870~\mum\ continuum (black contours, as in panel `a'). 
Blueshifted emission has been integrated from 4.0 to 6.1~\kms, while redshifted emission has been integrated from 9.7 to 11.1~\kms. 
Blue contours range from 21 to 99\% of the peak intensity (42.9~\jpb\,\kms), increasing in steps of 15\%.
Red contours range from 15 to 99\% of the peak intensity (33.8~\jpb\,\kms), increasing in steps of 5\%.
The colorscale is the first-order moment of the \ceo\,(2--1) emission showing a velocity gradient perpendicular to the outflow.
{\bf c)} Zoom in on \srce: SMA \tco\,(2--1) emission (blue contours correspond to the 9.0~\kms\ velocity, and red contours correspond to the 12.9~\kms\ velocity) and VLA 2.1~cm continuum emission (black contours, Rodr\'iguez et al. 2014).  
Blue contours range from 37 to 99\% of the peak intensity (1.43 \jpb\,\kms), increasing in steps of 15\%.
Red contours range from 37 to 99\% of the peak intensity (0.89 \jpb\,\kms), increasing in steps of 10\%.
Black contours (2.1 cm) range from $-3$, 3, and 6 times the rms noise of the map, 4.7~$\mu$\jpb.
The colorscale is the Spitzer/IRAC2 image showing a brightness increase at the position of the blueshifted lobe. 
}
\label{flargefov}
\end{center}
\end{figure*}

\subsection{Ancillary data}
We searched the Canada France Hawaii Telescope (CFHT), {\it Spitzer} and {\it Herschel} archives for images of the region and found that it was observed at several epochs in the optical and near-infrared (CFHT) as well as with {\it Spitzer} (Joergensen et al. 2006; Rebull et al. 2007; Evans et al. 2009), and {\it Herschel}. Upper limits in the $i$, $z$, $J$, $H$, $K_\mathrm{s}$ bands were estimated by adding scaled Point Spread Functions (PSFs) of decreasing fluxes at the expected position of the source until the detection disappeared. The PSF used were taken from the images themselves, using a nearby point source with high signal-to-noise.

\subsubsection{CFHT {\it MegaCam} }
IC348-SMM2 was observed with {\it MegaCam} in the $i$ and $z$ bands in the course of programs 09BH43, 06BF29, 07BH50, and 13BD96. The individual pipeline processed images were retrieved from the Canadian Astronomy Data Center (CADC) archive. Individual exposures ranging from 3 to 300~s were obtained in each filter. The whole set of images adds up to a total exposure time of 2\,990~s and 9\,201~s in the $i$ and $z$-band, respectively. The individual frames were astrometrically and photometrically registered using {\sc Scamp} (Bertin 2006) and stacked using {\sc SWarp} (Bertin et al. 2002).

\subsubsection{CFHT {\it WIRCam}}
IC348-SMM2 was observed with {\it WIRCam} in the $J$, $H$, and $K_\mathrm{s}$ broad-band filter and H$_2$ narrow-band filter. The individual pipeline processed and sky-subtracted images were retrieved from the CADC archive. Individual exposures ranging from 5 to 200~s were obtained.  The whole set of images adds up to a total exposure time of 3\,710~s, 1\,360~s, 13\,455~s and 24\,810~s in the $J$, $H$, $K_\mathrm{s}$ and H$_2$ bands, respectively.  The individual frames were astrometrically and photometrically registered using {\sc Scamp} and stacked using {\sc SWarp}. 

\begin{table*}
\caption{Parameters of the 870~\mum\ continuum sources detected with the SMA}
\begin{center}
{\small
\begin{tabular}{lcccccccc}
\noalign{\smallskip}
\hline\noalign{\smallskip}
&\multicolumn{2}{c}{Position$^\mathrm{a}$}
&Dec. ang. size~$^\mathrm{a}$
&Physical size~$^\mathrm{a}$
&Dec. P.A.~$^\mathrm{a}$
&$I_\mathrm{\nu}^\mathrm{peak}$~$^\mathrm{a}$
&$S_\mathrm{\nu}$~$^\mathrm{a}$
&Mass$^\mathrm{b}$
\\
\cline{2-3}
Source
&$\alpha (\rm J2000)$
&$\delta (\rm J2000)$
&$(''\times'')$
&(AU$\times$AU)
&($^\circ$)
&(\jpb)
&(Jy)
&(\mj)\\
\noalign{\smallskip}
\hline\noalign{\smallskip}
SMM2   	&03:43:57.06 	&32:03:04.6 &\phb$2.20\times1.74$ 	&$530\times420$	&$75.8$	&$0.610\pm0.017$	&$0.97\pm0.19$      	&250\\
SMM2E	&03:43:57.73  	&32:03:10.1 &$<1.4\times1.1$			&$<340\times260$	&-    		&$0.113\pm0.017$	&$0.11\pm0.02$	&29\\
\hline
\end{tabular}
\begin{list}{}{}
\item[$^\mathrm{a}$] Position, deconvolved size, peak intensity, and flux density are derived by fitting a Gaussian in the image domain.
Uncertainty in the peak intensity is the rms noise of the cleaned image, $\sigma$.  
Uncertainty in flux density has been calculated as $\sqrt{(\sigma\,\theta_\mathrm{source}/\theta_\mathrm{beam})^{2}+(\sigma_\mathrm{abs})^{2})}$ (Beltr\'an \et\ 2001), where $\theta_\mathrm{source}$ and $\theta_\mathrm{beam}$ are the size of the source and the beam respectively, and $\sigma_\mathrm{abs}$ is the error in the absolute flux scale, which takes into account the uncertainty on the calibration applied to the flux density of the source ($S_\nu\times\%_\mathrm{uncertainty}$).
\item[$^\mathrm{b}$] Masses derived assuming a dust temperature of 24~K (Section 4), and a dust (and gas) mass opacity coefficient of 0.0175~cm$^2$\,g$^{-1}$ (obtained by interpolating the tabulated values of Ossenkopf \& Henning 1994, see Section~\ref{srescont}). The uncertainty in the masses due to the opacity law is estimated to be a factor of 2 (Ossenkopf \& Henning 1994).
\end{list}
}
\end{center}
\label{tcont}
\end{table*}

\subsubsection{{\it Herschel Space Observatory}}

The Perseus molecular clouds were observed by the {\it Herschel Space Observatory} as part of the Gould Belt Survey (Andr\'e et al. 2010). A first set of observations was obtained in parallel mode using the PACS (70, 100, and 160~$\mu$m) and SPIRE (250, 350, and 500~$\mu$m) instruments simultaneously, but in this work we present the 100~\mum\ data only from the Gould's Belt project. More details about the observational strategy can be found in Andr\'e et al. (2010). 
The \emph{Herschel} images at 70 and 160~\mum\ were obtained from the  program GT2\_zbalog\_2 (P.I. Balog, integration time of 15792~s), for which observations were performed in scan mode.
The data were pre-processed using the {\it Herschel} Interactive Processing Environment (Ott 2010) version 10.0.2843, and the latest available version of the calibration files. The final maps were subsequently produced using Scanamorphos version 21 (Roussel 2013), using its galactic option, as recommended to preserve large scale extended emission.

\begin{figure*}
\begin{center}
\begin{tabular}[b]{cc}
    \epsfig{file=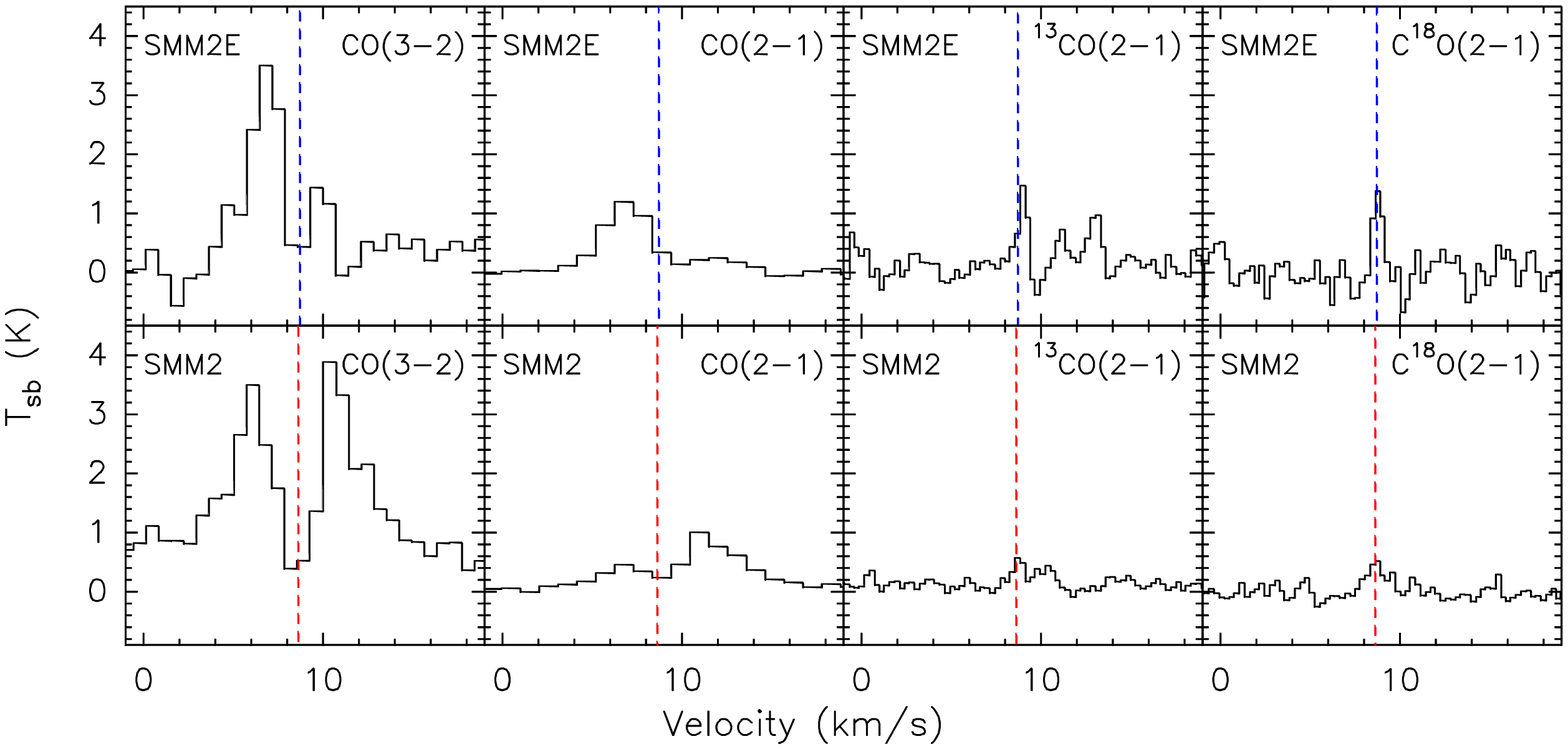, width=7.cm,angle=270}\\
\end{tabular}
\caption{
\emph{Top, from left to right:} CO\,(3--2), CO\,(2--1), \tco\,(2--1), and \ceo\,(2--1) spectra for \srce, averaged over a polygon of $\sim8$~arcsec of diameter centred on the source.
\emph{Bottom, from left to right:} idem for \src, averaging over a polygon of $8\times14$~arcsec$^2$.
Vertical dashed lines indicate the $v_\mathrm{lsr}$ for \srce\ ($\sim8.72$~\kms) and for \src\ ($\sim8.63$~\kms) measured in the \ceo\ spectra.
}
\label{fspecs}
\end{center}
\end{figure*}

\section{Results}

\subsection{Continuum}\label{srescont}

In Fig.~\ref{flargefov}a we present the SMA 870~\mum\ continuum emission (contours), overlaid on the H$_2$ 2.12~\mum\ image from the CFHT archive (Section 2.3.2), showing the large-scale H$_2$ knots of the HH\,797 outflow. 
The dominating submillimetre source in the field, SMM2 (Walawender et al. 2006), is the driving source of the large-scale outflow (\eg\ Pech et al. 2012). 
In addition to \src, we detect 870~\mum\ emission from \srce\ at about 10~arcsec to the northeast ($\sim2400$~AU). The source is detected up to 6 times the rms noise level and is unresolved (see zoom in Fig.~\ref{flargefov}b). 
This 870~\mum\ source has a counterpart at 1.3~mm (detected with the SMA, Chen \et\ 2013, who labelled the source `MMS2') and at 2.1~cm (detected with the Jansky Very Large Array (JVLA) by Rodr\'iguez et al. 2014, and labelled as `JVLA3c').

Since \srce\ is unresolved by the SMA at 870~\mum, we adopt a flux density equal to the peak intensity, 0.11~Jy. Assuming a dust temperature of 24~K (see Section~4), and a dust mass opacity coefficient at 870~\mum\ of 1.751 cm$^2$\,g$^{-1}$ (column 6 of Table~1 of Ossenkopf \& Henning 1994, corresponding to agglomerated dust grains with thin ice mantles at densities $\sim10^6$~cm$^{-3}$), the flux density measured with the SMA corresponds to a total mass of gas and dust of $\sim30$~\mj. We estimated an uncertainty in the masses due to uncertainty in the dust opacity of about a factor of 2 (Ossenkopf \& Henning 1994).
%
The positions, deconvolved sizes, peak intensities, flux densities and mass estimates of both \src\ and \srce\ are listed in Table~\ref{tcont}.

\begin{table*}
\caption{Physical parameters of the CO\,(3--2) outflow driven by \srce}
\centering
\footnotesize
\begin{tabular}{lccc cccc cc}
\hline\hline\noalign{\smallskip}
&$t_\mathrm{dyn}$
&dec. size
&$N_{12}$~$^\mathrm{a}$
&$M_\mathrm{out}$~$^\mathrm{a}$
&$\dot{M}$~$^\mathrm{a}$
&$P$~$^\mathrm{a}$
&$\dot{P}$~$^\mathrm{a}$
&$E_\mathrm{kin}$~$^\mathrm{a}$
&$L_\mathrm{mech}$~$^\mathrm{a}$
\\
Lobe
&(yr)
&(arcsec)
&(cm$^{-2}$)
&(\mo)
&(\mo~yr$^{-1}$)
&(\mo~\kms)
&(\mo~\kms~yr$^{-1}$)
&(erg)
&(\lo)
\\
\noalign{\smallskip}
\hline\noalign{\smallskip}
Red	&670	&$<1.4\times1.1$	&$1.1\times10^{16}$&$5.3\times10^{-6}$&$7.8\times10^{-9}$&$7.4\times10^{-6}$&$1.1\times10^{-8}$&$1.0\times10^{38}$	&$7.4\times10^{-7}$\\
Blue	&530	&$2.2\times1.1$	&$3.8\times10^{16}$&$2.8\times10^{-5}$&$5.3\times10^{-8}$&$5.9\times10^{-5}$&$1.1\times10^{-7}$&$1.2\times10^{39}$	&$8.6\times10^{-6}$\\
All	&600	&$-$				&$4.9\times10^{16}$&$3.3\times10^{-5}$&$5.6\times10^{-8}$&$6.7\times10^{-5}$&$1.1\times10^{-7}$&$1.3\times10^{39}$	&$9.3\times10^{-6}$\\
\hline
\hline
\end{tabular}
\begin{list}{}{}
\item[$^\mathrm{a}$] Parameters are calculated following Palau \et\ (2007, 2013), without correcting for inclination. 
From the line peak of the CO\,(3--2) spectrum, $\sim3.5$~K, we estimated an excitation temperature of $\sim10$~K, assuming optically thick emission in the line centre.
The CO\,(3--2) outflow parameters were corrected for opacity using a correction factor of 10, taken as a first approach from the facts that the opacity of CO\,(2--1) is measured to be $\sim18$ (see Section~\ref{sresco32}), and that CO\,(3--2) is typically optically thinner than CO\,(2--1).
\end{list}
\label{toutfpar}
\end{table*}

\subsection{Molecular line data}\label{sresco32}

Spectra of the average emission inside a polygon of $\sim8$~arcsec of diameter ($8\times14$~arcsec$^2$) centred on \srce\ (\src) are shown in Fig.~\ref{fspecs} for CO\,(3--2), CO\,(2--1), \tco\,(2--1) and \ceo\,(2--1).
The CO\,(3--2) spectra for both \srce\ and \src\ show a double-peaked profile which could be due in part to the filtering of emission at systemic velocities and/or self-absorption of the cold foreground cloud. However, while the blueshifted and redshifted wings have similar intensities for \src, this is not the case of \srce, for which the peak of emission is found at 6.8--7.5~\kms. For these velocities, the emission of \srce\ dominates over the emission of the \src\ outflow (within one primary beam), and consists of one main lobe elongated roughly in the east-west direction. It also presents some deep negative features which probably arise from large-scale emission filtered out by the SMA (see channel maps in Fig.~A1 of the Appendix). This suggests that the emission at these velocities traces part of the large-scale ambient cloud. 
The CO\,(2--1) and \tco\,(2--1) spectra towards \srce\ indicate that the CO\,(2--1) emission is optically thick (opacity $\sim 18$), as the intensity of the CO\,(2--1) is only a factor of 4 larger than the intensity of \tco\,(2--1) (once smoothed to the same spectral resolution, and assuming a $^{12}$CO/$^{13}$CO ratio of 65, Chin et al. 1995). For this reason, we refrain from studying the outflow using CO\,(2--1), and rather present the emission traced by \tco\,(2--1) at two particular velocities (systemic, and redshifted), showing one lobe to the north(east) and another fainter lobe to the south(west) of \srce\  (see Fig.~\ref{flargefov}c).
On the other hand, the \ceo\,(2--1) line from \srce\ is narrow (0.67~\kms\ of line width), and can be used to infer the systemic velocity of \srce, which is found to be very similar to the systemic velocity of \src\ (8.72 and 8.63~\kms, respectively, after gaussian fitting of the spectra).
Concerning the morphology of the \ceo\,(2--1) emission associated with \srce, shown in Fig.~\ref{flargefov}b, it is compact and presents a velocity gradient in the (south)east-(north)west direction (see also the channel maps in Fig.~\ref{fc18o21ch} of the Appendix).

Fig.~\ref{flargefov}b presents the zero-order moments of the CO\,(3--2) emission, revealing a blueshifted compact lobe to the north(east) and a redshifted compact lobe to the south(west) of \srce, similar to what is seen in \tco\,(2--1) (Fig.~\ref{flargefov}c). Blueshifted emission in \srce\ is stronger than redshifted emission, and ranges from 4 to 6.5~\kms, while the redshifted emission ranges from 9.7 to 11~\kms. 
The blueshifted lobe is separated from the redshifted lobe by 3--4~arcsec, or 700--1000~AU, and the blueshifted lobe size is 2--3~arcsec ($\sim600$~AU) while the red lobe is unresolved. Furthermore, the blueshifted lobe matches a 4.5~\mum\ brightness increment as seen with {\it Spitzer} (see Fig.~\ref{flargefov}c). The redshifted lobe is more compact and fainter than the blueshifted one. This difference, which has been seen in other cases (\eg\ Pety et al. 2006; Fern\'andez-L\'opez et al. 2013) and is predicted by simulations (\eg\ Offner et al. 2011), could be due to the fact that the redshifted emission is located towards the southwest, where the extinction (and opacity) increases because of the extended envelope of SMM2.

We derived the CO\,(3--2) column density (from the 8~arcsec averaged spectrum, Fig.~\ref{fspecs}) following Palau \et\ (2007, 2013).
To do this, we first estimated the excitation temperature assuming optically thick emission in the line centre (see Table~\ref{toutfpar}), and calculated the line area for the velocity ranges where the outflow wings are detected (given in Fig.~\ref{flargefov}). 
We used an opacity correction factor for CO\,(3--2) of $\sim10$, adopted to be smaller than the measured opacity for CO\,(2--1) ($\sim18$, see above).
The resulting CO column density is $\sim 5\times10^{16}$~cm$^{-2}$.
As for the mass, we used the (deconvolved) sizes given in Table~\ref{toutfpar}, adopted a mean molecular weight of 2.8, and a CO abundance of $X$(\co)=$10^{-4}$ (\eg\ Frerking, Langer, \& Wilson 1982),
obtaining a final (blueshifted + redshifted lobes) mass of $3\times10^{-5}$~\mo. 
The outflow parameters given in Table~\ref{toutfpar} were estimated also following Palau et al. (2007), and without correcting for inclination. 
We obtained a mass outflow rate of $6\times10^{-8}$~\mo\,yr$^{-1}$, an outflow force of $10^{-7}$~\mo\,\kms\,yr$^{-1}$, and a mechanical luminosity of $\sim10^{-5}$~\lo\ (Table~\ref{toutfpar}).

\begin{table}
\caption{Photometry for \srce}
\begin{center}
{\small
\begin{tabular}{ccccl}
\noalign{\smallskip}
\hline\noalign{\smallskip}
$\lambda$
&$S_\nu$
&$\sigma_\mathrm{abs}$\supa
&Beam
\\
($\mu$m)
&(mJy)
&(mJy)
&(arcsec)
&Instrument
\\
\noalign{\smallskip}
\hline\noalign{\smallskip}
0.75		&$<0.00049$  	&$-$			&$-$			&CFHT/Megacam\\
0.88		&$<0.00062$  	&$-$			&$-$			&CFHT/Megacam\\
1.25		&$<0.0098$  	&$-$			&$-$			&CFHT/WIRCAM\\
1.62		&$<0.021$  	&$-$			&$-$			&CFHT/WIRCAM\\
2.13		&$<0.016$  	&$-$			&$-$			&CFHT/WIRCAM\\
3.6\supb	&$<0.0026$  	&$-$			&1.7			&Spitzer/IRAC\\
4.5\supb	&0.021  		&0.006		&1.7			&Spitzer/IRAC\\
5.8\supb	&0.10  		&0.04		&1.9			&Spitzer/IRAC\\
8.0\supb	&0.16  		&0.10		&2.0			&Spitzer/IRAC\\
24\supb	&3.2  		&0.4			&6.0			&Spitzer/MIPS\\
70		&360 	 	&50			&5.6			&Herschel/PACS\\
100		&1200  		&100		&6.8			&Herschel/PACS\\
160		&1300  		&700		&11			&Herschel/PACS\\
350   	&1400    		&600        		&10			&CSO\\
450   	&1000     		&400         	&9			&JCMT\\
850   	&110		&20			&$2.8\times2.3$	&SMA\\
1300\supb&65 			&15			&$3.3\times3.0$	&SMA\\
21000\supb&0.027 		&0.008		&$2.4\times1.7$	&JVLA\\
33000\supb&$<0.016$ 	&$-$			&$3.9\times2.5$	&JVLA\\
\hline
\end{tabular}
\begin{list}{}{}
\item[$^\mathrm{a}$] Absolute flux uncertainty as described in Table~\ref{tcont}. 
\item[$^\mathrm{b}$] The measurements of Spitzer/IRAC+MIPS, SMA at 1.3~mm, and JVLA at 2.1 and 3.3~cm are reported in Evans et al. (2009), Chen et al. (2013) and Rodr\'iguez et al. (2014).
\end{list}
}
\end{center}
\label{tsed}
\end{table}

\begin{figure*}
\begin{center}
\begin{tabular}[b]{c}
    \epsfig{file=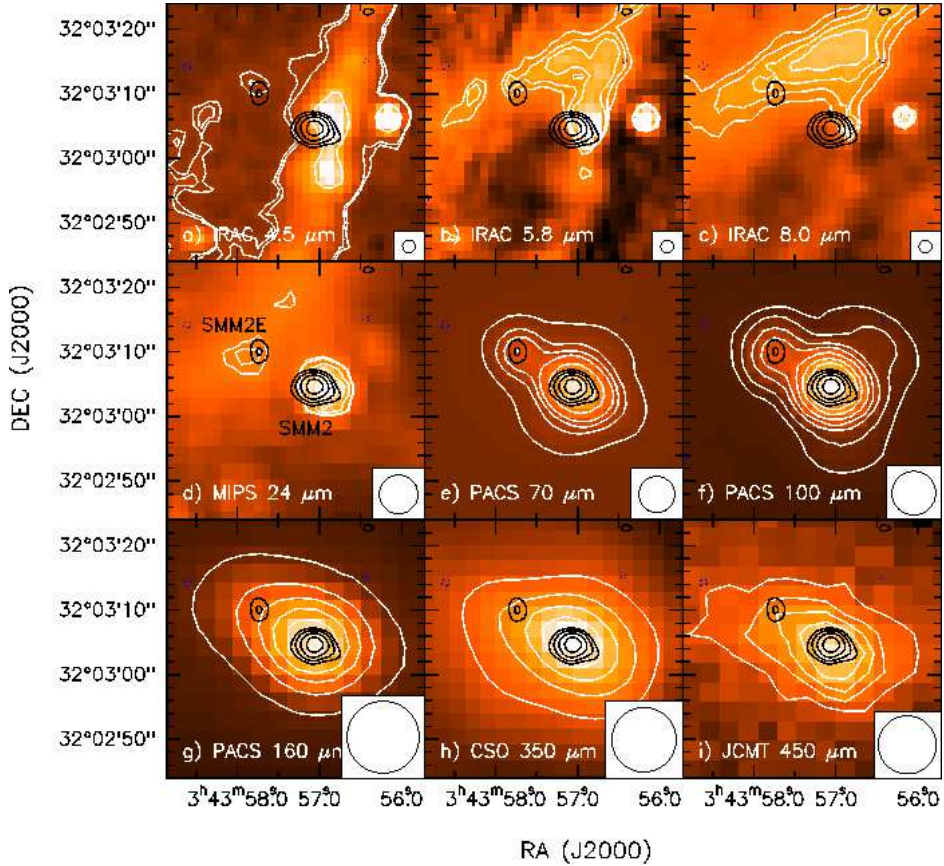, scale=0.7,angle=0}\\
\end{tabular}
\caption{Color scale and white contours: emission at different wavelengths from 4.5 to 450~\mum\ in the field of view of \src. Contours are: 
panel `a' (IRAC 4.5~\mum): 0.48, 0.50, 1.1, and 1.8~MJy\,sr$^{-1}$;
panel `b' (IRAC 5.8~\mum): 5.3, 5.4, and 5.5~MJy\,sr$^{-1}$;
panel `c' (IRAC 8.0~\mum): 17.0, 17.3, 17.6, and 17.9~MJy\,sr$^{-1}$;
panel `d' (MIPS 24~\mum): 43.4, 43.6, 44.5, 45.5, and 46.5~MJy\,sr$^{-1}$;
panel `e' (PACS 70~\mum): 0.15, 0.30, 0.45, 0.65, 1.2, 2.0, and 3.0~\jpb;
panel `f' (PACS 100~\mum): 1.0, 1.5, 2.5, 3.5, 5.0, 7.0, and 9.0~\jpb;
panel `g' (PACS 160~\mum): 3, 6, 9, 12, 15, and 18~\jpb;
panel `h' (CSO 350~\mum): 4.8, 5.9, 7.1, 8.3, and 9.5~\jpb;
panel `i' (JCMT 450~\mum): 0.25, 0.75, 1.25, 1.75, and 2.5~\jpb.
In all panels, black contours correspond to the SMA 870~\mum\ emission (contours as in Fig.~\ref{flargefov}), and the beam is shown in the bottom right corner of each panel. }
\label{fdiffwavel}
\end{center}
\end{figure*}

\section{Analysis}\label{sana}

\subsection{Building the SED for \srce}

With the aim of building the SED and estimating the bolometric luminosity of \srce, we used the archive data reported in Section 2.3. While the object remains undetected in the CFHT and Spitzer/IRAC1 bands, there are detections in IRAC2, IRAC3, IRAC4, MIPS1 (fluxes reported in Evans et al. 2009) and Herschel/PACS bands, which we present in Fig.~\ref{fdiffwavel}. It is worth noting that IRAC3, IRAC4 and MIPS1 images also reveal an elongated structure oriented in the southeast-northwest direction and passing through \srce\ (Fig.~\ref{fdiffwavel}b--d).

Regarding the \emph{Herschel}/PACS bands, at 70~\mum\ \srce\ is well separated from \src\ (Fig.~\ref{fdiffwavel}e). In order to estimate the flux of \srce\ at 70~\mum, we convolved the PSF of PACS at 70~\mum\ by a Gaussian\footnote{The Gaussian size for the convolution was estimated from a Gaussian fit to the 70~\mum\ emission in \src.} (to take into account the possible contribution of the PSF side lobes and the extended emission from \src\ to \srce), and subtracted the convolved PSF from the observed image. The residual image presents a clear excess of emission at the position of \srce, and we fitted such an excess with a Gaussian + constant level, providing us with the flux for \srce\ at this wavelength (Table~\ref{tsed}). Uncertainties were estimated by using different boxes for the Gaussian + constant level fit.
We estimated the fluxes of \srce\ at 100 and 160~\mum\ (Figs.~\ref{fdiffwavel}f,g) by applying the same technique.

In panel `h' of Fig.~\ref{fdiffwavel} we present the CSO 350~\mum\ continuum emission of the observations described in Section~2.2. The morphology of the emission is fully consistent with the results obtained by Davidson et al. (2011), with the difference that our image is calibrated in absolute flux scale. We fitted a Gaussian + constant level in a region including only the western part of SMM2, and the residual image presents an excess at the position of SMM2E, of 1.4~\jpb.
At slightly longer wavelengths, we used the 450~\mum\ image of Walawender \et\ (2006, see Fig.~\ref{fdiffwavel}i) to fit a Gaussian + constant level as in the CSO image. The residual shows an excess of emission up to 1.0~\jpb\ at the position of \srce, and we adopt this value to build the SED. Finally, we also used the values measured with the SMA at 870~\mum\ (this work), 1.3~mm (Chen et al. 2013), and the value at 2.1~cm (Rodr\'iguez et al. 2014). In Table~\ref{tsed} we list the measured fluxes (or upper limits) for \srce\ at different wavelengths.

\begin{figure}
\begin{center}
\begin{tabular}[b]{cc}
    \epsfig{file=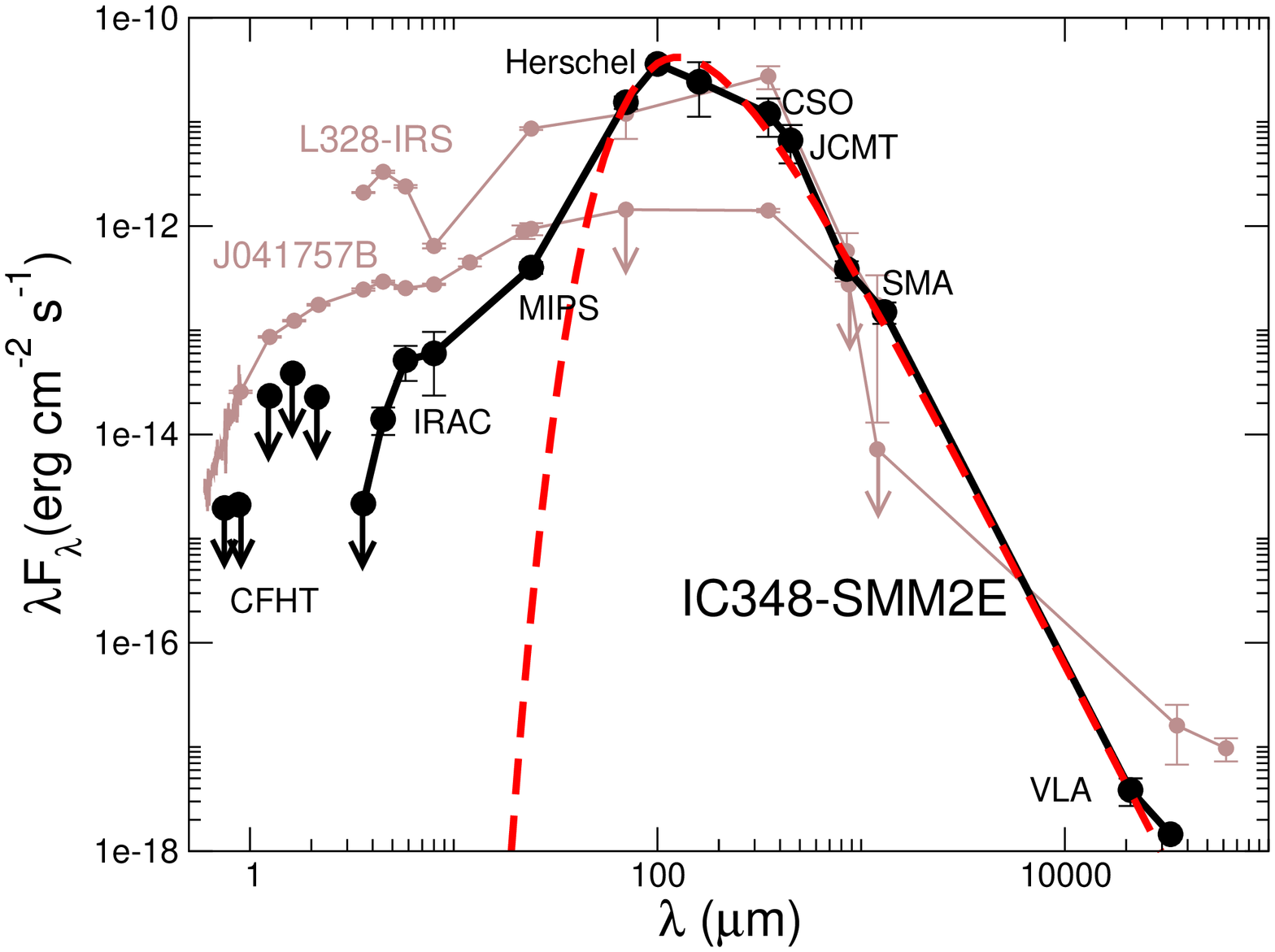, scale=0.4,angle=270}\\
\end{tabular}
\caption{SED of IC348-SMM2E (black thick solid line) compared to two previously reported proto-BD candidates (brown thin solid lines), J041757-B (Barrado et al. 2009; Palau et al. 2012) and L328-IRS (Lee et al. 2009, 2013). The SED suggests that \srce\ is a more embedded object than the other proto-BDs, being a SED typical of Class 0 low-mass protostars. The dashed (red) line shows the fit of a modified blackbody of 24~K, 35~\mj, and dust emissivity index of 0.8.
}
\label{fsed}
\end{center}
\end{figure} 

\begin{figure*}
\begin{center}
\begin{tabular}[b]{cc}
    \epsfig{file=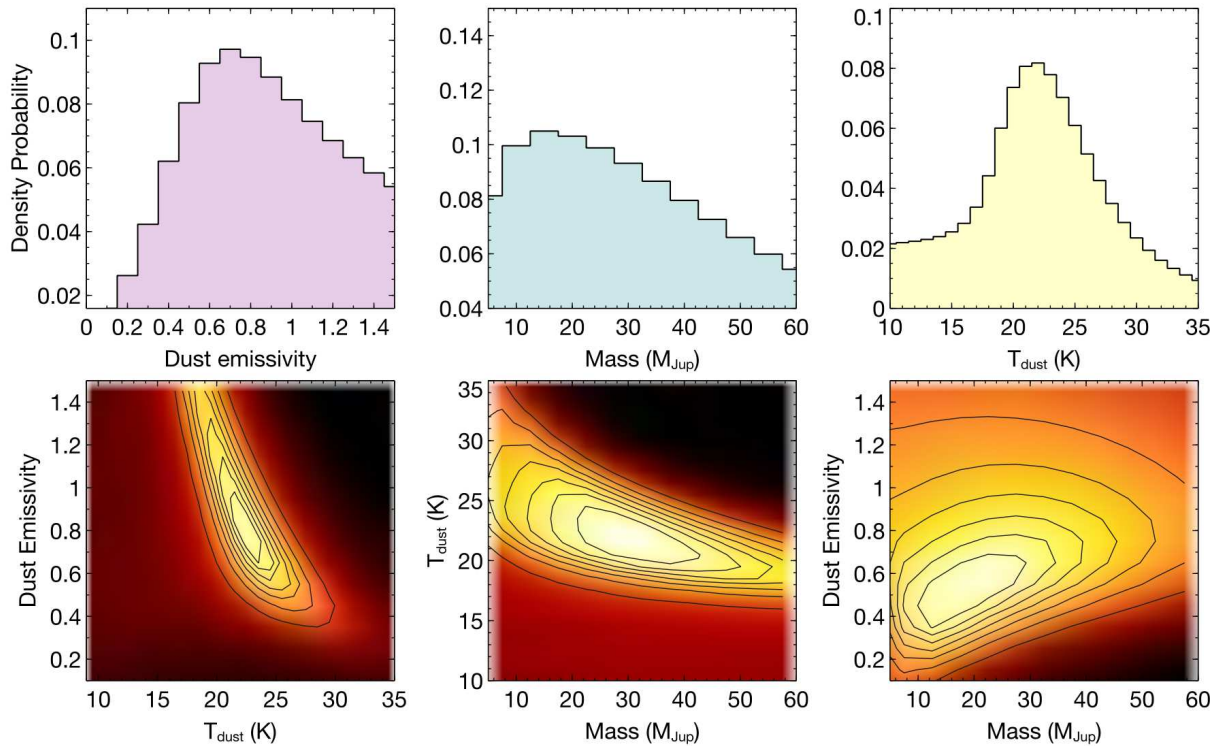, width=16cm,angle=0}\\
\end{tabular}
\caption{
{\emph Top:} Probability density functions of the three parameters of the modified blackbody fit (shown in Fig.~\ref{fsed} with a red dashed line). 
{\emph Bottom:} $\chi^{2}$ maps of the fit for the different pairs of parameters. 
}
\label{fbayes}
\end{center}
\end{figure*}

\subsection{SED properties}

By using the flux measurements estimated in the previous section we built the SED of \srce\ (Fig.~\ref{fsed}). The SED peaks near 100~\mum\ and decreases about two orders of magnitude at near-infrared wavelengths, with no flatness in the mid-infrared range. This is not seen in more evolved objects which have already been detected in the near-infrared, such as the proto-BD candidates J041757B (Barrado et al. 2009; Palau et al. 2012) and L328-IRS (Lee et al. 2009, 2013). The SEDs of these two objects are also shown in Fig.~\ref{fsed} for comparison, and both present brighter fluxes in the near/mid-infrared than \srce, while the submillimetre fluxes are comparable. 

A fit of a modified black-body to the SED of \srce\ (red dashed line in Fig.~\ref{fsed}) can account for all the measured fluxes from 70~\mum\ down to the centimetre range, within uncertainties. 
Thus, the SED of \srce\ is typical of `early' Class 0 young stellar objects (\eg\ Enoch et al. 2009). From these data we estimated a bolometric temperature for \srce\ of $\sim35$~K, which falls within the range of values of Class 0 objects ($T_\mathrm{bol}<70$~K; Chen et al. 1995). 
To do the fit of the modified black body, we used two methods. First, we searched for the parameters minimizing the $\chi^2$ in the ranges
10$\le T_{\rm d} \le$35~K (by steps of 1~K), 5$\le M_\mathrm{env} \le$60~M$_{\rm Jup}$ (by steps of 5~\mj), and 0.1$\le \beta \le$1.5 (by steps of 0.1), and found the minimum $\chi^2$ for a dust temperature of 24~K, an envelope mass of 35~\mj, and a dust emissivity index of 0.8 (adopting the same opacity at 870~\mum\ as in Section~3, and assuming that the opacity changes with frequency following a power law of index $\beta$). 
Second, we performed a Bayesian analysis by comparing the SED of the target with the grid of 2\,574 modified blackbody models covering the ranges given above. For each model, the reduced $\chi^{2}$ and corresponding relative probability $\mathrm{e}^{-\chi^{2}/2}$ were computed. The relative likelihood for each of the parameters was obtained by a marginalization of the parameter space successively over each dimension. Figure~\ref{fbayes} shows the corresponding probability density functions, from which we derived the most probable values. Uncertainties on these values were estimated using bootstrapping and computing the 2.5\% and 97.5\% confidence interval over 1000 replications. The Bayesian analysis yields that the most probable values of the parameters are: 22 +2/$-2$~K, 15 +15/$-5$~\mj, and 0.7 +0.3/$-0.1$, for the dust temperature, envelope mass and dust emissivity index, respectively. These values are in  good agreement with the values minimizing $\chi^{2}$, with the envelope mass slightly shifted towards lower masses.

\section{Discussion}

\subsection{Will \srce\ remain substellar?}

From the SED presented in the previous section, we estimate a bolometric luminosity for \srce\ of 0.10~\lo.
This bolometric luminosity already classifies \srce\ as a VeLLO.
In addition, \srce\ is not likely a young stellar object in a quiescent phase because the mechanical luminosity estimated from the outflow is $\sim10^{-5}$~\lo\ (Section~\ref{sresco32}), much smaller than the bolometric luminosity.

The measured bolometric luminosity is an upper limit to the internal luminosity of the object, because part of the luminosity might be produced by the interstellar radiation field. A first approach to the internal luminosity of an object can be obtained if the flux at 70~\mum\ is known (Dunham et al. 2008). Using equation (2) of Dunham et al. (2008), and for the flux measured by us at 70~\mum\ (Section 4) we obtain an internal luminosity for \srce\ of $0.06$~\lo, corresponding to a mass of $\sim75$~\mj\ according to the evolutionary models of Baraffe et al. (2002, Fig.2), and for the earliest possible time computed for these models (1~Myr).
The accreted mass of the central object is probably smaller than 75~\mj\ because for Class 0 objects the envelope mass is comparable to the accreted mass (\eg\ Andr\'e et al. 1993; Dunham et al. 2008), and for the envelope mass we estimated a value $\la35$~\mj\ (Sections~3 and 4). In addition, part of the internal luminosity might come from accretion.

An independent estimate of the stellar mass can be made assuming that the internal luminosity arises entirely from accretion (a reasonable assumption for Class 0 objects).
Assuming the Shu (1977) model, the mass accretion rate at which an initially isothermal core accretes, $\dot{M}_\mathrm{acc}$, depends only on the sound speed $c_\mathrm{s}$:
$\dot{M}_\mathrm{acc} = \frac{\eta_\mathrm{\dot{M}}}{G}\, c_\mathrm{s}^3$, where $G$ is the gravitational constant and $\eta_\mathrm{\dot{M}}$ is the efficiency in the mass accretion rate. Thus, $\dot{M}_\mathrm{acc}$ can be written in terms of the initial temperature of the core as $\dot{M}_\mathrm{acc} = \frac{\eta_\mathrm{\dot{M}}}{G}\, \Big( \frac{k T}{\mu m_\mathrm{H}}\Big)^{3/2}$.

On the other hand, the accretion luminosity is directly proportional to the mass accretion rate:

\begin{equation}\label{eLacc}
L_\mathrm{acc} = \eta_\mathrm{L}\,\frac{G m_* \dot{M}_\mathrm{acc}}{R_*},
\end{equation}

\noindent where $\eta_\mathrm{L}$ is the accretion luminosity efficiency with respect to steady spherical infall (for steady accretion through an optically thick disk,  $\eta_\mathrm{L}\sim1/2$, Hartmann 1998), and $m_*$ and $R_*$ are the mass and radius of the central hydrostatic object.  Thus, $L_\mathrm{acc}$ and $m_*$ are related linearly and $m_*$ can be written in the form:

\begin{equation}\label{emacc}
m_* = L_\mathrm{acc}\frac{R_*}{\eta_\mathrm{L}\,\eta_\mathrm{\dot{M}}} \Big( \frac{\mu m_\mathrm{H}}{k T}\Big)^{3/2}.
\end{equation}



We assumed $\eta_\mathrm{\dot{M}}\sim0.1$ (Terebey et al. 2006; Myers 2014), and an initial gas temperature $T$ around 10 K, which corresponds to a mass accretion rate of $\sim1.6\times10^{-7}$~\mo\,yr$^{-1}$, only a factor of two larger than the mass outflow rate inferred for \srce\ (see Table~\ref{toutfpar}). For the radius $R_*$, we adopted a range from 0.1~\ro\ (for pre-main sequence brown dwarfs, Stassun et al. 2012; Sorahana et al. 2013) to 1~\ro\ (from the simulations following the collapse of a core to stellar densities, yielding about $<3$~\ro\ for the second hydrostatic core for a collapsing core of 1~\mo, Bate, Tricco \& Price 2014). For this range of values, and assuming that the internal luminosity comes from accretion (to be conservative), we obtain that $m_*$ ranges from 2 to 24~\mj, well within the brown dwarf domain.
This accreted mass is fully consistent with the dynamical mass estimated from the \ceo\,(2--1) velocity gradient shown in Fig.~\ref{flargefov}b, of 16~\mj. The velocity gradient is seen perpendicular to the outflow and thus could be tracing rotation of an envelope/disk of about 300~AU of radius.
If confirmed, this would be the first disk found around a Class 0 proto-BD.



The reservoir of mass still available in the envelope is $\la35$~\mj\ (Section 4.2). 
This is comparable to the mass which should be accreted in the future assuming that the object will continue accreting at the same rate during the typical timescale of the Class 0 phase, (5--10)$\times10^4$~yr (e.g., Froebrich et al. 2006; Evans et al. 2009; Enoch et al. 2009; Dunham \& Vorobyov 2012). 
Thus, the final mass should be the sum of the current mass (2 to 24~\mj) plus the mass which will be accreted in the future ($\la35$~\mj), which corresponds to a final mass $<59$~\mj. Therefore, the true final mass of \srce\ will most likely remain substellar.

\begin{figure}
\begin{center}
\begin{tabular}[b]{cc}
    \epsfig{file=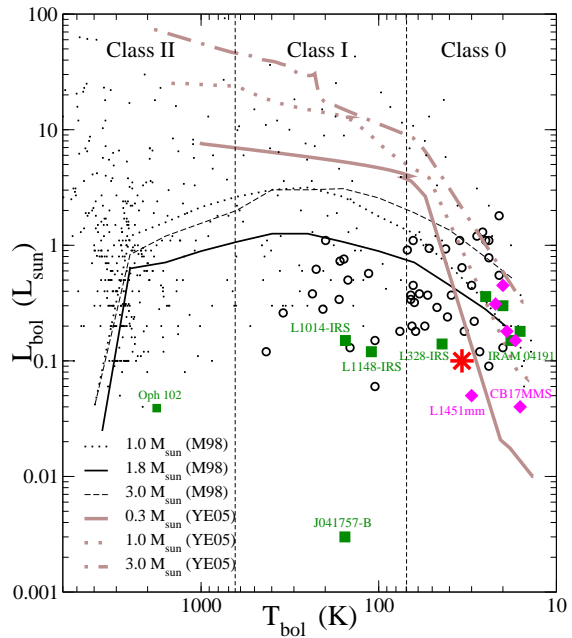, scale=0.55,angle=0}\\
\end{tabular}
\caption{Bolometric luminosity vs. bolometric temperature diagram after
Young \& Evans (2005), Dunham et al. (2008) and Barrado et al. (2009). The 50 sources
compiled by Dunham et al. (2008), showing some evidence for being
embedded low luminosity sources, are shown as empty black circles.
The thick brown lines correspond to the evolutionary tracks for the
three models with different masses (in solar masses) considered by
Young \& Evans. (2005). The thin black lines show the evolutionary
tracks for three models considered by Myers et al. (1998).
The dots correspond to Taurus members and the vertical dashed
lines show the Class 0/I and Class I/II $T_\mathrm{bol}$  boundaries from Chen
et al. (1995). VeLLOs are marked as green squares, FHCs are marked as magenta diamonds, and \srce\ is shown as a large red asterisk (see Table~\ref{tvellos}).
}
\label{flboltbol}
\end{center}
\end{figure}

\subsection{Comparison of \srce\ with other proto-BD candidates and young stellar objects}

Since the discovery of VeLLOs (Young et al. 2004), a number of objects with bolometric luminosities $\la0.5$~\lo\ have been studied in detail and have been found to be driving outflows. In Table~4 we compile the main properties of the most studied VeLLOs and FHCs. However, in most of the cases the available mass reservoir is large enough to cast doubt on the final substellar nature of the objects. In addition, their SEDs present different shapes suggestive of different evolutionary stages. 

To visualize this more clearly, in Fig.~\ref{flboltbol} we plot the position of VeLLOs (green squares) and FHCs (magenta diamonds) in a $L_\mathrm{bol}$ vs $T_\mathrm{bol}$ diagram. The figure shows that \srce\ is the least luminous of the Class 0 VeLLOs, and in Table~\ref{tvellos} we show that it is associated also with the least massive envelope. Only L328-IRS (Lee et al. 2009, 2013) has $L_\mathrm{bol}$, $T_\mathrm{bol}$, and envelope mass comparable to those in \srce. These two objects are very similar, not only in the properties of their SEDs, but also in the velocity extension of their outflows, of $\sim2$~\kms\ wide. The main difference between \srce\ and L328-IRS is the outflow size. The \srce\ outflow has not been detected in single-dish observations, and the interferometric images reveal lobes of only $\sim500$~AU (Table~\ref{toutfpar}), while the outflow driven by L328-IRS extends up to $\sim20000$~AU, indicating a longer lifetime. The fact that the mid-infrared fluxes of \srce\ are much lower than the L328-IRS fluxes (Fig.~\ref{fsed}) also indicates that \srce\ is probably younger than L328-IRS.
Regarding the FHCs, there are only two with lower luminosity than that of \srce, L1451mm (Pineda et al. 2011), and CB17-MMS (Chen et al. 2012). The mass reservoir (estimated from single-dish) of CB17-MMS is of the order of $\sim1$~\mo\ (Launhardt et al. 2010), being easy for CB17-MMS to achieve stellar masses. 
On the other hand, the outflow lobes of L1451mm are very compact, $\sim500$~AU, and the outflow velocities cover only $\la2$~\kms\ (Pineda et al. 2011), comparable to those in \srce. However, the mass reservoir of L1451mm is five times larger than that of \srce\ (Table~\ref{tvellos}).
%
Thus, \srce\ turns to be among the youngest Class 0 proto-BD candidates with a substellar-mass envelope.

\begin{figure}
\begin{center}
\begin{tabular}[b]{c}
    \epsfig{file=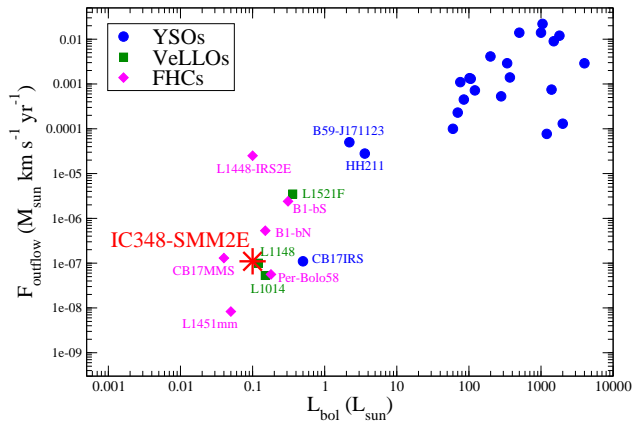, width=7.cm,angle=270}\\
\end{tabular}
\caption{Comparison of outflow force vs bolometric luminosity for objects observed with mm/submm interferometers. Blue dots correspond to young stellar objects (YSOs) from Beltran et al. (2008), Palau \et\ (2006), Chen et al. (2012), Hara et al. (2013) and Duarte-Cabral et al. (2013). Green squares correspond to Very Low Luminosity Objects (VeLLOs), and magenta diamonds correspond to First Hydrostatic Cores (FHCs, see references in Table~\ref{tvellos}). The red asterisk corresponds to \srce\ (this work).
}
\label{fcorrel}
\end{center}
\end{figure} 

\begin{table*}
\caption{Summary properties of VeLLOs and FHCs}
\begin{center}
{\small
\begin{tabular}{lccc cccl}
\noalign{\smallskip}
\hline\noalign{\smallskip}
&$T_\mathrm{bol}$
&$L_\mathrm{bol}$
&$L_\mathrm{int}$
&$M_\mathrm{env}$\supa
&$F_\mathrm{out}^\mathrm{SD}$\,\supb
&$F_\mathrm{out}^\mathrm{interf}$\,\supb
\\
Source
&(K)
&(\lo)
&(\lo)
&(\mo)
&(\mo\ \kms\ yr$^{-1}$)
&(\mo\ \kms\ yr$^{-1}$)
&Refs.\supc
\\
\noalign{\smallskip}
\hline\noalign{\smallskip}
VeLLOs\\
\noalign{\smallskip}
\hline\noalign{\smallskip}
IRAM04191	&18	&0.15	&0.07	&0.60	&$1.5\times10^{-5}$		&$-$				&1\\
L1521F-IRS	&25	&0.36	&0.04	&0.87	&$-$					&$3.5\times10^{-6}$	&2, 3, {\bf 4}\\
L1148-IRS	&110&0.12	&0.10	&0.14	&$-$					&$1.0\times10^{-7}$	&5\\
L673-7-IRS	&16	&0.18	&0.04	&0.39	&$1.0\times10^{-6}$		&$-$				&6, 7\\
L1014-IRS	&154&0.15	&0.09	&0.36	&$-$					&$5.3\times10^{-8}$	&8, 9\\
L328-IRS		&44	&0.14	&0.05	&0.09	&$2.0\times10^{-7}$		&$-$				&10, 11\\
GF9-2		&20	&0.30	&$<0.3$	&$-$		&$6.1\times10^{-8}$		&$-$				&12, 13\\
J041757B		&155&0.003	&$<0.003$&$0.01$	&$-$					&$-$				&14, 15\\
IC348-SMM2E	&35	&0.10	&0.06	&0.03	&$-$					&$1.1\times10^{-7}$&16\\
\noalign{\smallskip}
\hline\noalign{\smallskip}
FHCs\\
\noalign{\smallskip}
\hline\noalign{\smallskip}
Cha-MMS1	&20	&0.45	&0.15	&0.80	&$-$					&$-$				&17, 18\\
Per-Bolo58	&19	&0.18	&0.012	&1.00	&$-$					&$5.6\times10^{-8}$	&19, 20\\
L1448-IRS2E	&$-$	&$<0.1$	&$<0.1$	&0.50	&$-$					&$2.5\times10^{-5}$	&21\\
L1451-mm	&30	&0.05	&$<0.03$	&0.15	&$-$					&$8.3\times10^{-9}$	&22\\
CB17-MMS	&$<16$&$<0.04$&$<0.04$&4.0	&$-$					&$1.3\times10^{-7}$	&23, 24\\
B1-bS		&22	&0.31	&0.1--0.2	&7.3		&$-$					&$2.4\times10^{-6}$	&25, 26, 27\\
B1-bN		&17	&0.15	&$<0.03$	&9.4		&$-$					&$5.3\times10^{-7}$	&25, 26, 27\\
\hline
\end{tabular}
\begin{list}{}{}
\item[$^\mathrm{a}$] Envelope masses taken from Kauffmann et al. (2011) when available (being thus the mass measured with a single-dish within a radius of 4200~AU). For objects not included in the compilation of Kauffmann et al. (2011) the envelope masses are those measured with single-dish as reported in the literature. 
\item[$^\mathrm{b}$] $F_\mathrm{out}^\mathrm{SD}$ and $F_\mathrm{out}^\mathrm{interf}$ refer to the outflow force as observed with a single-dish telescope and an interferometer, respectively.
\item[$^\mathrm{c}$] Refs: (1) Andr\'e et al. (1999); (2) Bourke et al. (2006); (3) Takahashi et al. (2013); (4) Onishi, Mizuno \& Fukui (1999); (5) Kauffmann et al. (2011); (6) Dunham et al. (2010); (7) Schwarz et al. (2012); (8) Bourke et al. (2005); (9) Maheswar et al. (2011); (10) Lee et al. (2009); (11) Lee et al. (2013); (12) Wiesemeyer et al. (1999); (13) Furuya et al. (2006); (14) Barrado et al. (2009); (15) Palau et al. (2012); (16) this work; (17) Belloche et al. (2006); (18) Tsitali et al. (2013); (19) Enoch et al. (2010); (20) Dunham et al. (2011); (21) Chen et al. (2010); (22) Pineda et al. (2011); (23) Launhardt et al. (2010); (24) Chen et al. (2012); (25) Pezzuto et al. (2012); (26) Huang \& Hirano (2013); (27) Hirano \& Liu (2014).
\end{list}
}
\end{center}
\label{tvellos}
\end{table*}

There are a number of well-known correlations established for large samples of young stellar objects. If BDs form as a scaled-down version of low-mass stars, we expect that the properties of proto-BDs should fit also in these correlations. In particular, the outflow force seems to correlate with the bolometric luminosity of the driving source (\eg\ Bontemps \et\ 1996; Wu et al. 2004; Takahashi \& Ho 2012). In Fig.~\ref{fcorrel} we plot the outflow force of \srce\ and its bolometric luminosity on data from the literature of outflows observed with interferometers (using similar configurations), in order to avoid estimates of the missing flux which are necessarily uncertain. For the low and intermediate-mass young stellar objects, we used the compilation by Beltran \et\ (2008), and added the works of Palau et al. (2006), Chen et al. (2012), Hara et al. (2013), and Duarte-Cabral et al. (2013). For the lower mass objects we mainly included the VeLLOs and FHCs whose outflow has been detected with an interferometer\footnote{We did not include low-luminosity objects whose SED resembles those of Class II YSOs, likely corresponding to later evolutionary stages compared to the objects we are discussing here. This is the case of Oph-102 (Phan-Bao et al. 2008) and MHO5 (Phan-Bao et al. 2011), among others.} 
(L1521F: Bourke et al. 2006, Takahashi et al. 2013; L1148-IRS: Kauffmann et al. 2011; L1014-IRS: Bourke et al. 2005, Maheswar et al. 2011;
L1448-IRS2E: Chen et al. 2010; Per-Bolo58: Enoch et al. 2010, Dunham et al. 2011; L1451mm: Pineda et al. 2011; CB17-MMS and CB17-IRS: Chen et al. 2012).
The figure shows that \srce\ falls on the expected position in this diagram if one extrapolates the trend traced by the sample of young stellar objects (blue dots).
%
%
Thus, \srce\ seems to behave as a scaled-down version of low-mass stars, and probably will keep its mass substellar,  constituting one of the few examples of an excellent Class 0 proto-BD candidate.


\subsection{Is \srce\ forming isolated? A possible wide triple system}

In the previous section we showed that \srce\ seems to be a proto-BD candidate forming as a scaled-down version of low-mass stars. Something which remains to be answered however is whether \srce\ is forming isolated or is a companion of \src.
In the 5.8, 8.0 and 24~\mum\ images there is an extended structure elongated in the southeast-northwest direction and \srce\ lies in its southern tip (see Fig.~\ref{fdiffwavel}b,c,d). This structure is reminiscent of the `striations' seen near filaments in both high-mass and low-mass star-forming clouds (\eg\ Goldsmith et al. 2008; Busquet et al. 2013; Palmeirim et al. 2013).
However, \srce\ could just fall in projection at the tip of the striation. In addition, this striation is not seen in any of the Herschel bands, suggesting it must be relatively warm, and such a warm structure is not expected to fragment into substellar mass fragments, making this possibility unlikely.


Another possibility is that  \srce\ is gravitationally bound to \src. \src\ is a Class 0 young stellar object for which a bolometric luminosity of 0.4--1.1~\lo\ is measured (\eg\ Hatchell et al. 2007\footnote{We recalculated the value of the bolometric luminosity given by Hatchell \et\ (2007) to the IC\,348 distance adopted in this paper.}; Enoch et al. 2009). If we assume that this bolometric luminosity comes from accretion, and use equation (2) (for a radius similar to the solar radius), we obtain an accreted mass for \src\ of 0.2--0.4~\mo.
On the other hand, the difference of peak (radial) velocities of \srce\ and \src\ seen in the \ceo\,(2--1) spectra is around $\sim0.1$~\kms, which is consistent with a mass for \src\ of $\la0.4$~\mo\ if \srce\ is orbiting around \src.
Thus it seems plausible that both objects constitute a multiple system. Actually, Rodr\'iguez et al. (2014) show that \src\ seems to be a binary itself, making this group of objects a possible triple system, with very wide separations of 720 and 2400~AU (Chen et al. 2013; Rodr\'iguez et al. 2014). If this scenario is confirmed, we would be witnessing the formation of a very fragile triple system, of mass ratio $\sim0.1$ and widest separation of 2400~AU (corresponding to a binding energy of $9\times10^{40}$~erg). 
Whether such a fragile system will be disrupted in the future remains an open question, as typical timescales for BD ejection estimated from numerical simulations are around 0.1--0.2~Myr (\eg\ Basu \& Vorobyov 2012; Bate 2012), comparable to the lifetime of Class 0 objects. 

The number of known substellar wide (400--4000~AU) binaries is small (\eg\  Luhman et al. 2009; Radigan et al. 2009; Aller \et\ 2013; Bonavita \et\ 2014), and the \srce\ + \src\ system constitutes, to the best of our knowledge, the youngest substellar wide multiple system known to date, indicating that the fragmentation process that formed this wide system took place at the earliest stages of star formation.
The discovery of the Class 0 proto-BD \srce, forming as a scaled-down version of low-mass protostars, and belonging to a wide multiple system, strongly suggests that for this particular case the star formation processes extend down to substellar masses.

\section{Conclusions}

We present observations carried out with the SMA of the 870~\mum\ continuum and CO\,(3--2), \tco\,(2--1), and \ceo\,(2--1) emission of IC\,348-SMM2E, a faint millimetre source lying near the HH\,797 outflow in the IC\,348 cluster. We complement these data with archive data from CFHT, {\it Spitzer}, {\it Herschel}, and JCMT, and with new CSO observations, allowing us to characterize the main properties of such a faint object. Our main conclusions are summarized as follows:

\begin{itemize}

\item We detect two sources at 870~\mum, one strong and resolved, called SMM2 and which is the driving source of the HH\,797 outflow, and another much fainter, unresolved, and located about 10~arcsec to the (north)east of SMM2, called SMM2E. We estimate a mass of gas and dust of 250~\mj\ for SMM2 and 30~\mj\ for SMM2E. SMM2E is coincident with recently reported sources at 1.3~mm and 2.1~cm (Chen et al. 2013; Rodr\'iguez et al. 2014).

\item The SMA CO\,(3--2) emission reveals, in addition to the well-known HH\,797 outflow, a compact bipolar low-velocity outflow associated with SMM2E. The outflow wings cover 1--2~\kms, as found in other VeLLOs, and we estimate an outflow mass of $3\times10^{-5}$~\mo, a mass outflow rate of $6\times10^{-8}$~\mo\,yr$^{-1}$, and an outflow force of $10^{-7}$~\mo\,\kms\,yr$^{-1}$. An estimate of the mass accretion rate assuming the Shu (1977) model yields $\sim1.6\times10^{-7}$~\mo\,yr$^{-1}$.

\item \ceo\,(2--1) emission is detected towards \srce, which shows a velocity gradient perpendicular to the outflow, and corresponds to a dynamical mass of 16~\mj. In addition, the \ceo\ emission allows us to measure the systemic velocities of both \srce\ and \src, finding them to be very similar, which suggests \srce\ is gravitationally bound to \src\ and thus forms a wide (2400~AU) multiple system.

\item The analysis of {\it Spitzer}, {\it Herschel}, JCMT, and CSO data consistently presents hints of excess of emission at the position of SMM2E. By estimating the fluxes from 70 to 450~\mum, and using our collected data and data from the literature, we built the SED for SMM2E, and found that it can be fitted from 70~\mum\ to 2.1~cm with one single modified black-body with a dust temperature of 24~K, a dust emissivity index of 0.8 and a mass envelope of $\sim35$~\mj. Thus, the SED of SMM2E is typical of Class 0 objects.

\item The bolometric temperature and luminosity of SMM2E are estimated to be $\sim35$~K and 0.10~\lo, and the internal luminosity is $\sim0.06$~\lo. These properties of SMM2E place the object among the least luminous and most embedded objects known so far, as compared to VeLLOs and FHCs. In addition, SMM2E also presents the smallest envelope mass (measured with single-dish) among the Class 0 VeLLOs and FHCs, strongly suggesting that its final mass will probably remain substellar. 

\item A comparison of the outflow force for SMM2E to those of other VeLLOs, FHCs, low-mass and intermediate-mass young stellar objects (all measured with an interferometer) shows that \srce\ matches the lower end of the known relation of outflow force vs bolometric luminosity for low-mass protostars, suggesting that in this case the formation of this proto-BD candidate can be explained as a scaled-down version of low-mass stars.

\end{itemize}

\section*{Acknowledgements}

We are deeply grateful to Helen Kirk, Josh Walawender, and Dough Johnstone for sharing the 450~\mum\ JCMT data, 
and to Jackie Davidson, and Giles Novak for sharing 350~\mum\ CSO data to compare to the new data acquired at the CSO.
The authors are also grateful to the referee for her/his comments that significantly improved the paper.
A.P., L.A.Z., and L.F.R. acknowledge the financial support from UNAM, and CONACyT, M\'exico.
H.B. is funded by the Spanish Ram\'on y Cajal fellowship program number RYC-2009-04497. 
This research has been funded by Spanish grants AYA2010-21161-C02-02, and AYA2012-38897-C02-01.
D. Li acknowledges the support  from National Basic Research Program of China (973 program) No. 2012CB821800 and NSFC No. 11373038.
This research used the facilities of the Canadian Astronomy Data Centre operated by 
the National Research Council of Canada with the support of the Canadian Space Agency.
Based on observations obtained with MegaPrime/MegaCam, a joint project of CFHT and CEA/DAPNIA,
at the Canada-France-Hawaii Telescope (CFHT) which is operated by the National Research Council (NRC)
of Canada, the Institute National des Sciences de l'Univers of the Centre National de la Recherche 
Scientifique of France, and the University of Hawaii.
Based on observations obtained with WIRCam, a joint project of CFHT, Taiwan, Korea, Canada, France,
and the Canada-France-Hawaii Telescope (CFHT) which is operated by the National Research Council (NRC) 
of Canada, the Institute National des Sciences de l'Univers of the Centre National de la Recherche 
Scientifique of France, and the University of Hawaii.
This work is based in part on data obtained as part of the UKIRT Infrared Deep Sky Survey,
made use of VOSA, developed under the Spanish Virtual Observatory project supported from the Spanish MICINN through grant AyA2008-02156, and of the SIMBAD database, operated at CDS, Strasbourg, as well as Topcat (Taylor 2005).

{}

\begin{appendix}

\section{CO\,(3--2) and \ceo\,(2--1) channel maps}

\begin{figure}
\begin{center}
\begin{tabular}[b]{cc}
    \epsfig{file=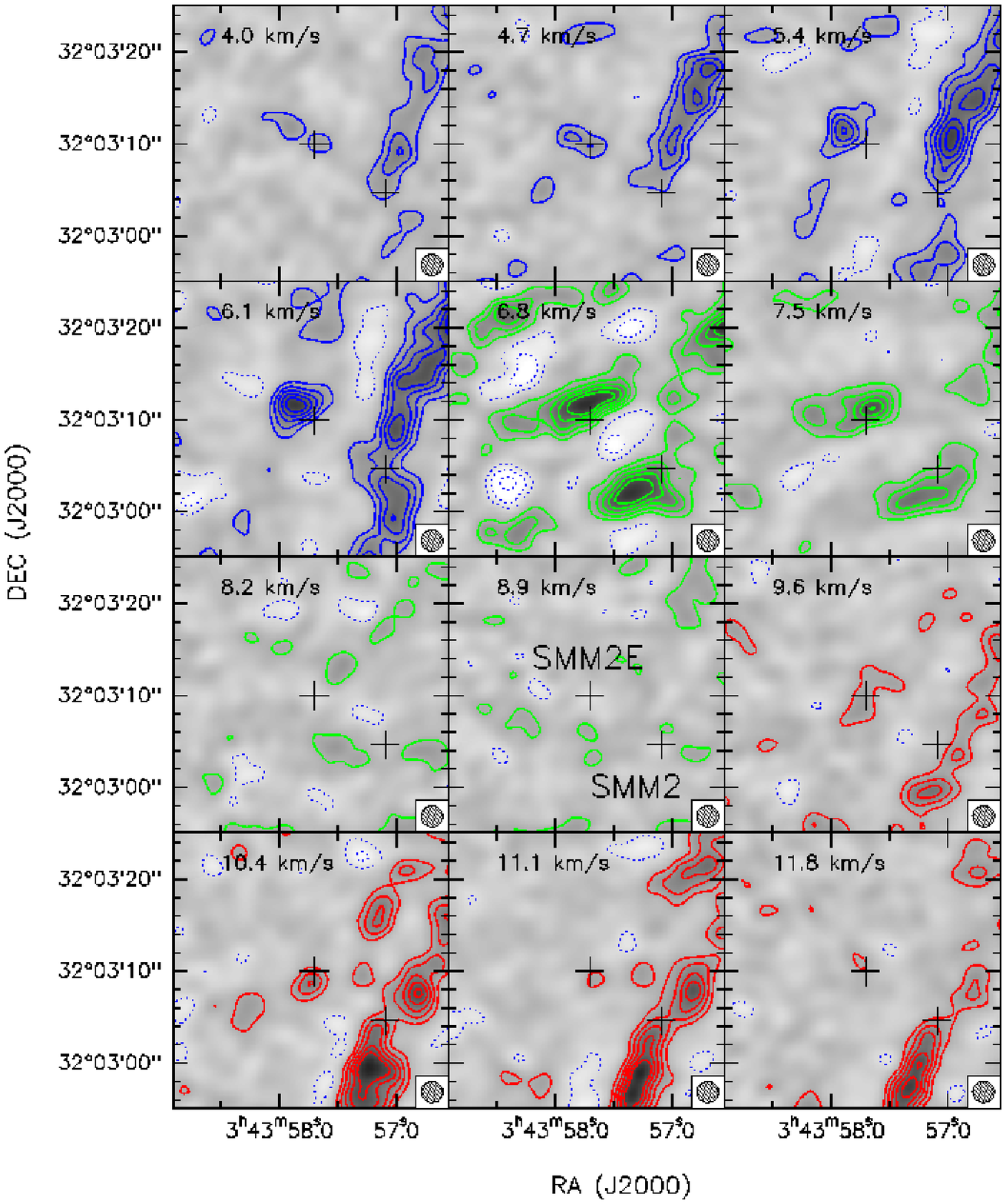, width=8.5cm,angle=0}\\
\end{tabular}
\caption{SMA CO\,(3--2) channel maps centred on \srce. The plus signs correspond to \srce\ (left) and \src\ (right). 
Contours are $-6$, $-4$, $-2$, 2, 4, 6, 8, and 10 times 0.85~\jpb. The velocity of each channel is indicated in the top-left corner, and the beam is shown in the bottom-right corner of each panel.
}
\label{fco32ch}
\end{center}
\end{figure}

\begin{figure}
\begin{center}
\begin{tabular}[b]{cc}
    \epsfig{file=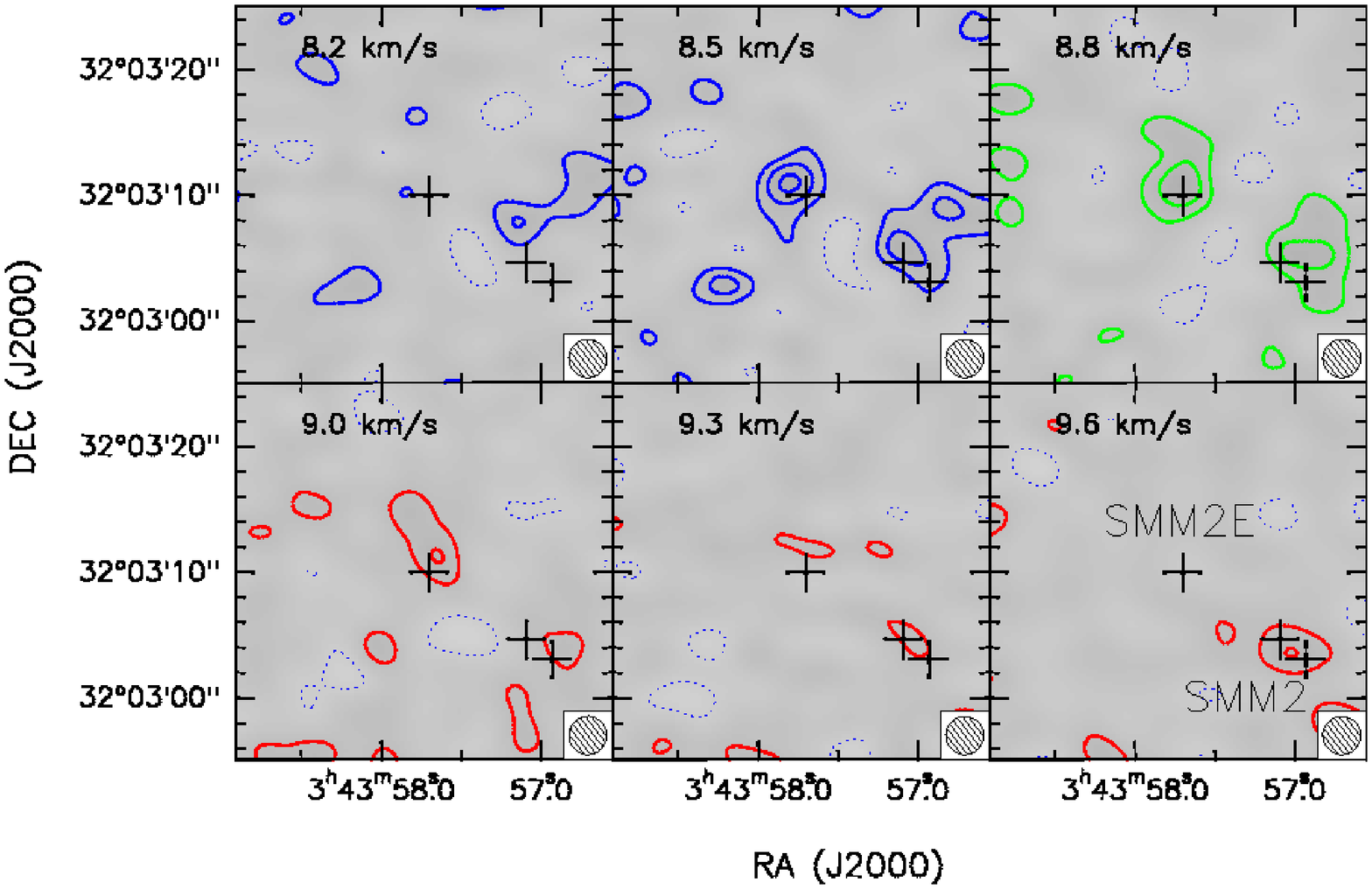, width=5.5cm,angle=270}\\
\end{tabular}
\caption{
SMA \ceo\,(2--1) channel maps centred on \srce. The plus signs correspond to \srce\ (left), \src-JVLA3b (centre), and  \src-JVLA3a  (right, Rodr\'iguez et al. 2014). 
Contours are $-6$, $-4$, $-2$, 2, 4, and 6 times 0.25~\jpb. The velocity of each channel is indicated in the top-left corner, and the beam is shown in the bottom-right corner of each panel.}
\label{fc18o21ch}
\end{center}
\end{figure}

In Figs.~\ref{fco32ch} and \ref{fc18o21ch} we present the channel maps of the CO\,(3--2) and \ceo\,(2--1) emission as observed with the SMA, and centred on \srce.
The emission associated with \src\ is part of the HH\,797 large-scale outflow (Pech et al., in prep.).

\end{appendix}

\end{document}